\documentclass[conference]{IEEEtran}

\usepackage{amsmath}
\DeclareMathOperator*{\argmin}{\arg\!\min}
\usepackage{epsfig}
\usepackage{epstopdf}
\usepackage{graphicx}
\usepackage{float}
\usepackage[noadjust]{cite}
\usepackage[]{algorithm2e}
\usepackage{mwe}
\usepackage{subcaption}

\begin{document}
 \title{ Log-Normal Matrix Completion for Large Scale Link Prediction}
 \author{Brian Mohtashemi
 \and
 Thomas Ketseoglou}
 \maketitle

\begin{abstract}
The ubiquitous proliferation of online social networks has led to the widescale emergence of relational graphs expressing unique patterns in link formation and descriptive user node features. Matrix Factorization and Completion have become popular methods for Link Prediction due to the low rank nature of mutual node friendship information, and the availability of parallel computer architectures for rapid matrix processing. Current Link Prediction literature has demonstrated vast performance improvement through the utilization of sparsity in addition to the low rank matrix assumption. However, the majority of research has introduced sparsity through the limited $L_1$ or Frobenius norms, instead of considering the more detailed distributions which led to the graph formation and relationship evolution. In particular, social networks have been found to express either Pareto, or more recently discovered, Log Normal distributions. Employing the convexity-inducing Lovasz Extension, we demonstrate how incorporating specific degree distribution information can lead to large scale improvements in Matrix Completion based Link prediction. We introduce Log-Normal Matrix Completion (LNMC), and solve the complex optimization problem by employing Alternating Direction Method of Multipliers. Using data from three popular social networks, our experiments yield up to $5\%$ AUC increase over top-performing non-structured sparsity based methods.
\end{abstract}

\section{Introduction}
As a result of widespread research on large scale relational data, the matrix completion problem has emerged as a topic of interest in collaborative filtering, link prediction \cite{15,16,17,18,19,20,21,22,23,24,25,26,27,28,29,30}, and machine learning communities. Relationships between products, people, and organizations, have been found to generate low rank sparse matrices, with a broad distribution of rank and sparsity patterns. More specifically, the node degrees in these networks exhibit well known Probability Mass Functions (PMFs), whose parameters can be determined via Maximum Likelihood Estimation. In collaborative filtering or link prediction applications, row and column degrees may be characterized by differing PMFs, which may be harnessed to provide improved estimation accuracy. Directed networks have unique in-degree and out-degree distributions, whereas undirected networks are symmetric and thus exhibit the same row-wise and column wise degree distributions. Though originally thought to follow strict Power Law Distributions, modern social networks have been found to exhibit Log Normal degree patterns in link formation \cite{2}.
\\
\hspace*{.5em} In this work, we propose Log Normal Matrix Completion (LNMC) as an alternative to typical $L_1$ or Frobenius norm constrained matrix completion for Link Prediction. The incorporation of the degree distribution prior generally leads to a non-convex optimization problem. However, by employing the Lovasz extension on the resulting objective, we reduce the problem to a convex minimization over the Lagrangian, which is subsequently solved with Proximal Descent and Alternating Direction Method of Multipliers (ADMM). Through experimentation on Google Plus, Flickr, and Blog Catalog social networks, we demonstrate the advantage of incorporating structured sparsity information in the resulting optimization problem.

\section{Related Work}
Link prediction has been thoroughly researched in the field of social network analysis as an essential element in forecasting future relationships, estimating unknown acquaintances, and deriving shared attributes. In particular, \cite{3} introduces the concept of the Social Attribute Network, and uses it to predict the formation and dissolution of links. Their method combines features from matrix factorization, Adamic Adar, and Random walk with Restart using logistic regression to give link probabilities. However, the calculations of such inputs may be time-intensive, and shared attributes may be unlikely, leading to non-descriptive feature vectors.
\\
\hspace*{.5em} Matrix Completion for Link Prediction has previously been investigated within the Positive Unlabeled (PU) Learning Framework, where the nuclear norm regularizes a weighted value-specific objective function \cite{1}. Although the weighted objective improves the prediction results, the subsequent optimization is non-convex and thus subject to instability. Binary Matrix completion employing proximal gradient descent is studied in \cite{7}, however, sparsity is not considered, and Link Prediction is not included in the experiment section. The structural constraints that must be satisfied for provably exact completion are described in \cite{8}. In this technical report, the required cardinality of uniformly selected elements is bounded based on the rank of the matrix. Unique rank bounds for matrix completion are considered in \cite{9}, where the Schatten p-Norm is utilized on the singluar values of the matrix. Matrix Completion for Power Law distributed samples is studied in \cite{11}, where various models are compared, including the Random Graph, Chung Lu-Vu, Preferential Attachment, and Forest Fire models. However, link prediction is not considered and the resulting optimization problem is non-convex.
\\
\hspace*{.5em} The concept of simultaneously sparse and low rank matrices was introduced in \cite{4}, where Incremental Proximal Descent is employed to sequentially minimize the objective, and threshold the singular values and matrix entries. Due to the sequentiality of the optimization, the memory footprint is reduced, however, the objective is non-convex and may result in a local minimum solution.  Also, the tested methods employed in simulation are elementary, and more advanced techniques are well known in the link prediction community. Simultaneous row and column-wise sparsity is discussed in \cite{11}, where a Laplacian based norm is employed on rows and a Dirichlet semi-norm is utilized on columns. A comparison between nuclear and graph based norms is additionally provided. In \cite{12}, Kim et. al present a matrix factorization method which utilizes group wise sparsity, to enable specifically targeted regularization. However, the datasets which we utilize do not identify group membership, and thus we will not consider affiliation in our prediction models.
\\
\hspace*{.5em} Structured sparsity was thoroughly investigated in \cite{5}, and applied to Graphical Model Learning. However, the paper focuses solely on the Pareto Distribution which characterizes scale-free networks, and does not cover the Log Normal Methods which are presented in this paper. Also, Link Prediction is not considered in the experimental section. Node specific degree priors are introduced in \cite{6}, and the Lovasz Extension is additionally employed to learn scale free networks commonly formed by Gaussian Models. However, the stability of the edge rank updating is not proven, and Log Normally distributed networks are not considered.
\\
\hspace*{.5em} The Lovasz Extension and background theory are presented in \cite{13}, where Bach provides an overview on submodular functions and minimization.

\section{Proposed Approach}
\subsection{Link Prediction}
In this paper, we consider social network graphs, since they have been proven to follow Pareto, and more recently discovered, Log Normal, degree distributions. The Social Network Link Prediction problem involves estimating the link status, $X_{i,j}$, between node $i$ and node $j$, where $X_{i,j}$ is limited to binary outcomes. Together, the set of all nodes, $V$, and links, $E$, form the graph $G=(V,E)$, where $E$ is only partially known. Unknown link statuses may exist when either the relationship between $i$ and $j$ is non-public, or the observation is considered unreliable over several crawls of the social network. Combined, the observations can be expressed in the form of a partial adjacency matrix, $A_{\Omega}$, which contains all known values in the set of observed pairs, $\Omega$. Unmeasured states between two nodes are set to 0 in $A_{\Omega}$. This matrix can be stored in sparse format for memory conservation, and operation complexity reduction.
\subsection{Structured Sparsity based Matrix Completion for Link Prediction}
As demonstrated in \cite{1,4,7}, Matrix Completion involves solving for unknown entries in matrices by employing the low-rank assumption in addition to other side information regarding matrix formation and evolution. Traditionally, matrix completion problems are expressed as
\begin{equation}
\hat{X} = \argmin_X  \|A_{\Omega} - X_{\Omega}\|_F^2 + \lambda \|X\|_{*},
\end{equation}
where
\begin{align*}
X_{\Omega \hspace*{.1em} {i,j}}=
\begin{cases}
X_{i,j}, \text{$if$ $\{i,j\}$ $\epsilon$ $\Omega$} \\
0, \hspace*{1.2em} otherwise, \\
\end{cases}
\end{align*}
$\|\cdot\|_{F}$ is the Frobenius norm, and $\|\cdot\|_{*}$ is the nuclear norm (Schatten p-norm with $p=1$). The nuclear norm can be defined as
\begin{equation}
\|X\|_{*} = \sum_{i=1}^{min\{m,n\}} \sigma_i,
\end{equation}
where $\sigma_i$ is the $i_{th}$ eigenvalue, when arranged in decreasing order, and $m$ and $n$ are the row count and column count, respectively. In this paper, $m$ is assumed equal to $n$. $\hat{X}$ is the estimated complete matrix after convergence is attained. Generally, these problems are solved using proximal gradient descent, which employs singular value thresholding on each iteration \cite{31}. However, this problem generally lacks incorporation of prior sparsity information encoded into the matrix. Thus we augment the problem as
\begin{equation}
\hat{X} = \argmin_X \|A_{\Omega} - X_{\Omega}\|_F^2 + \lambda_1 \|X\|_{*} + G(X),
\end{equation}
where $G$ is defined as follows:
\begin{equation}
G(X) = \lambda_2 \Gamma_{i,\alpha}(X) + \lambda_3 \Gamma_{j,\beta}(X).
\end{equation}
Here, $\Gamma_{i,\alpha}(X)$ is a sparsity inducing term, where $i$ implies that the sparsity is applied on matrix rows, j implies sparsity is applied on matrix columns, $\alpha$ is the prior in-degree distribution, and $\beta$ is the out-degree distribution.
For the rest of this paper, we will consider the case of symmetric adjacency matrices, and thus set $\lambda_3$ to $0$.
\subsection{Log-Normal Degree Prior}
As demonstrated in \cite{2}, many social networks, including Google+, tend to exhibit the Log-Normal Degree Distribution
\begin{equation}
p(d) = \frac{1}{d \sigma \sqrt{2\pi}} e^{-\frac{(\ln d - \mu)^2}{2\sigma^2}}.
\end{equation}
Thus we derive $\Gamma(X)$ as the Maximum Likelihood Estimate
\begin{equation}
\Gamma(X) = -\ln \prod _{i}p(d_{X_i}),
\end{equation}
where $d_{X_i}$ is the degree of the $i_{th}$ row of $X$, which simplifies to the following:
\begin{equation} \label{eq:-1}
\Gamma(X) = \sum_{i} \ln(d_{X_{i}}\sigma \sqrt{2 \pi}) + \frac{(\ln d_{X_{i}}-\mu)^2}{2\sigma^2}.
\end{equation}
This is equivalent to a summation of scaled Pareto Distributions with shape parameter 1 added to additional square terms. Thus the final optimization problem becomes
\begin{equation}\label{eq:zz}
\begin{split}
\hat{X} = \argmin_X \|A_{\Omega} - X_{\Omega}\|_F^2 + \lambda_1 \|X\|_{*} +\\
 \lambda_2 \sum_i \ln(d_{X_i}\sigma \sqrt{2 \pi}) + \frac{(\ln d_{X_i}-\mu)^2}{2\sigma^2}.
 \end{split}
\end{equation}
Due to the presence of the log term in the optimization, convex methods cannot be directly applied to the minimization, since the problem is not guaranteed to have an absolute minimum. Optimization of this problem is a multi-part minimization, which can be solved using the Alternating Direction Method of Multipliers (ADMM).

\subsection{Optimization}
ADMM allows the optimization problem to be split into less complex sub-problems, which can be solved using convex minimization techniques. In order to decouple (\ref{eq:zz}) into smaller subproblems, the additional variable, $Y$, is introduced as
\begin{gather*}
\argmin_X \|A_{\Omega} - X_{\Omega}\|_F^2 + \lambda_1 \|X\|_{*} + \Gamma(Y)\\
\text{s.t. } X=Y.
\end{gather*}
Expressing the problem in ADMM update form, the sequential optimization becomes
\begin{align}
&X^{k+1} = \argmin_X \hspace{.2em} \{ \|A_{\Omega} - X_{\Omega}\|_F^2 + \lambda_1 \|X\|_{*}\\ \nonumber& \hspace*{3.5em} +\frac{\mu}{2} \|X - Y^{k} + V^{k} \|_{F}^2 \} \\
&\label{eq:0}Y^{k+1} = \argmin_Y \lambda_2 \Gamma(Y) + \frac{\mu}{2} \|X^{k+1} - Y + V^{k} \|_F^2 \\
&V^{k+1} = V^{k} + X^{k+1} - Y^{k+1}.
\end{align}
In practice, step size values, $\mu$, in the range $[.01,.1]$ have been found to work well. Convergence is assumed, and the sequence is terminated once $\|X^{k+1} - X^{k}\|^2_F < \delta $. The initial values, $X^0$, $Y^0$ and $V^0$ are set to zeros matrices. Although ADMM has slow convergence properties, a relatively accurate solution can be attained in a few iterations. Due to the convexity of the initial equation, proximal gradient descent is employed for minimization.
The proximal gradient method minimizes problems of the form
\begin{equation}
\text{minimize } g(X) + h(X),
\end{equation}
using the gradient and proximal operator as
\begin{equation}\label{eq:a}
X^{k,l+1} = \text{prox}_{\psi^l h}(X^{k,l} - \psi^{l}\nabla g(X^{k,l})),
\end{equation}
where $\psi^{l+1} = \phi \psi^{l}$, and $\phi$ is a multiplier utilized on each gradient descent round. Typically a value of $.5$ is sufficient for $\phi$, leading to rapid convergence in $10$ rounds, however, a value $<.5$ would result in slower, but more accurate minimization. The optimal value for $\psi^0$ is determined through experimentation. For Log-Normal Matrix Completion, $g(X) = \|A_{\Omega} - X_{\Omega}\|_F^2 + \frac{\mu}{2} \|X-Y^k + V^k \|^{2}_F$, and $h(X) = \lambda \|X\|_{*}$. The proximal operator of $h(X)$ becomes a sequential thresholding on the eigenvalues, $\sigma$, of the argument in (\ref{eq:a})
\begin{equation}
\text{prox}_{\psi h} = Q \text{ diag }((\sigma_i - \psi)_{+})_i Q^T,
\end{equation}
where $Q$ is the matrix of eigenvectors. The subproblem reaches convergence when $\|X^{k,l+1} - X^{k,l}\|_F^2 < \kappa $.
The noise of the matrix is reduced through sequential thresholding, leaving only the strongest components of the low rank matrix. This algorithm is advantageous due to rapid convergence properties and automatic rank selection. Known as the Iterative Soft Thresholding Algorithm (ISTA), this method can be parallelized for gradient calculation and recombined for the Eigenvalue decomposition. Although the interim result of each round of minimization is generally not sparse, matrix entries with values below a given threshhold can be forced to 0 to allow sparse matrix Eigenvalue Decomposition (such as eigs in Matlab) to be performed with minimal error.
\subsection{Lovasz Extension}
  (\ref{eq:0}) is a non-convex optimization problem due to the log of the set cardinality function. However, the problem can be altered into a convex form using the Lovasz Extension on the submodular set function. As described in \cite{13}, the Lovasz Extension takes on the following form:
\begin{equation}
f(w) = \sum_{j=1}^n w_{z_j} [ F(\{z_1,... , z_j\}) - F(\{z_1,... , z_{j-1}\})].
\end{equation}
 Here, $z$ is a permutation of $j$ which ensures components of $w$ are ordered in decreasing fashion, $w_{z_1}\geq w_{z_2}\geq w_{z_n}$, and $F$ is a submodular set function. The Lovasz Extension is always convex when $F$ is submodular, thus allowing convex optimization techniques to be used on the resulting transformed problem.
 \\
In order to transform each individual row of sampled relationship information into a set, $S$, the support function, $S_i = \text{Supp }(X_i) $ is utilized. As a result $S_i \epsilon \{0,1\}^n$, where $n$ is the number of columns present in the matrix X. A submodular set function must obey the relationship
\begin{equation} \label{eq:1}
F(A \cup \{p\}) - F(A) \geq F(B \cup \{p\}) - F(B),
\end{equation}
where $A \subseteq B$, and ${p}$ is an additional set element. In this paper, $F$ is a log-normal transformation on the degree $d$. The degree, $d_i = \sum_i^n S_{i,j}$, is modular, and thus follows (\ref{eq:1}) with strict equality. Thus for $F$ to be sub modular, the subsequent transformation of the degree must be submodular as well.
\\
After applying the Lovasz Extension to (\ref{eq:-1}), the result is
\begin{align}\label{eq:3}
&\Gamma(X) = \sum_{i=1}^m \sum_{j=1}^n [\text{ln}^2 (j+1) - \text{ln}^2(j)\\ \nonumber &+ \frac{(\sigma^2-\mu)(\text{ln}(j+1) - \text{ln}(j))}{\sigma^2}]|X_{i,j}|.
\end{align}
Here, $|X|$ is used in order to maintain the positivity required for the Lovasz Extension to remain convex. Further details regarding the optimization of this problem can be obtained in Appendix A.
\subsection{Considerations}
In order for (\ref{eq:3}) to be utilized, (\ref{eq:-1}) must remain a submodular function of the degree. Thus, both the first derivative and the second derivative of the function must remain positive, creating the following constraint:
\begin{equation}
\text{ln}(d+\tau) \geq (1 + \mu - \sigma^2).
\end{equation}
 $\tau$ is introduced to prevent the left side of the inequality from approaching $-\infty$. In practice, a small constant is also subtracted or added from the obtained set function in order to assure that $F(\emptyset) = 0$. These small coefficients are determined during the Cross Validation phase, after obtaining the optimal $\sigma$ and $\mu$ values which satisfy the given constraints.
\section{Experiment}
In order to compare the performance of the LNMC method with other popular Link Prediction methods, an experiment was performed using several data sets from existing literature:
\begin{enumerate}
\item Google + - The Google + dataset \cite{3} contains $5,200$ nodes and  $24,690$ links, captured in AUG 2011. The data contains both Graph topology and node attribute information; however, the side-features are removed since our method requires edge status only.
\item Flickr -  Flickr is a social network based on image hosting, where users form communities and friendships based on common interests. The Flickr dataset \cite{14} contains $80,513$ nodes, $5,899,882$ links, and $195$ groups. Group affiliation was discarded due to irrelevance to the LNMC method.
\item Blog Catalog - Blog Catalog \cite{14} is a blogging site where users can form friendships, and acquire group membership. The utilized dataset contains $10,312$ nodes, $333,983$ links, and $39$ groups. Again, for the context of this paper, the group information was removed.
\end{enumerate}

\begin{figure}
        \begin{subfigure}[b]{0.175\textwidth}
                \includegraphics[trim = 14mm 0mm 0mm 0mm, width=30mm]{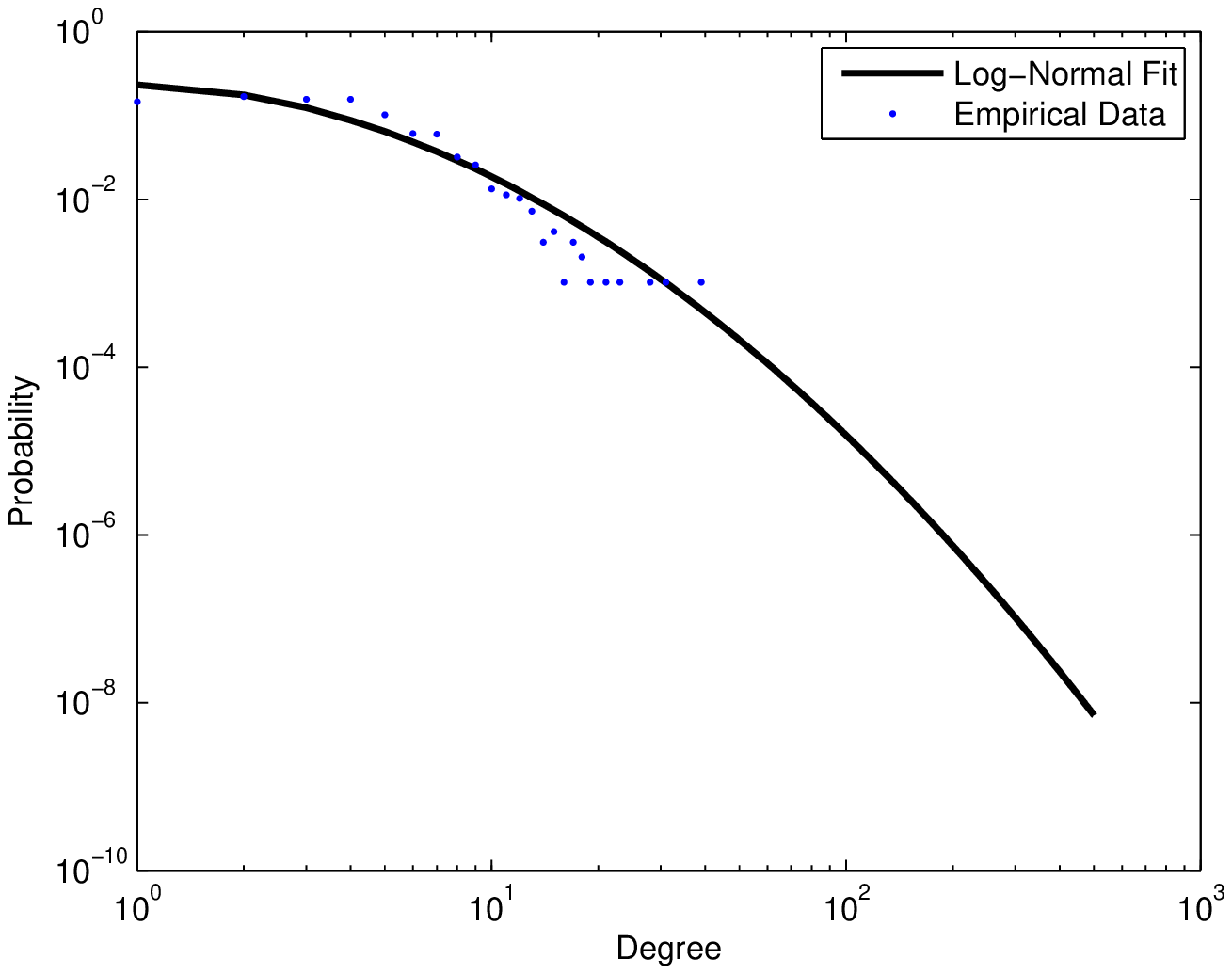}
                \centering
                \caption{Google+  }
                \label{fig:x}
        \end{subfigure}%
        \begin{subfigure}[b]{0.175\textwidth}
                \includegraphics[trim = 14mm 0mm 0mm 0mm, width=30mm]{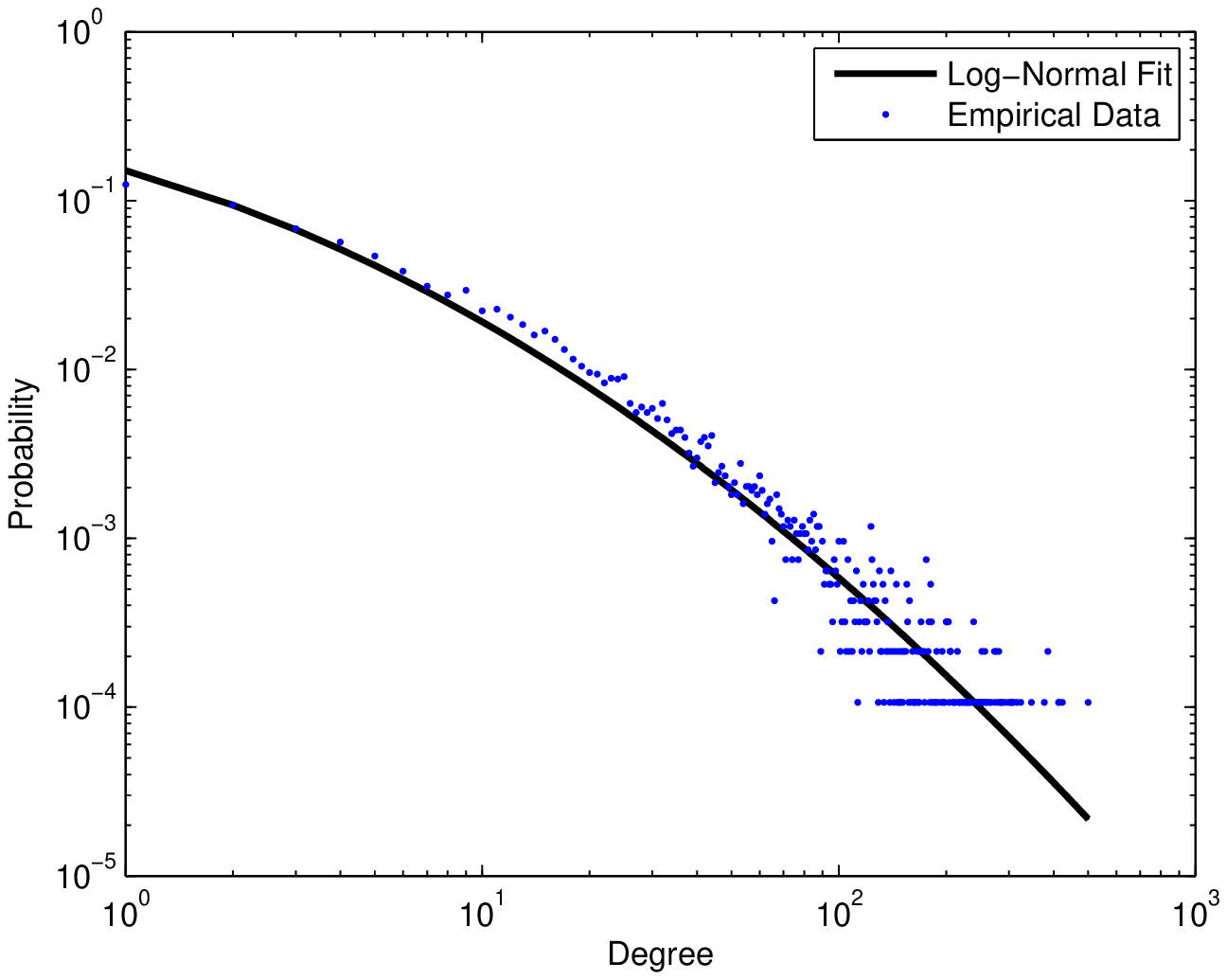}
                \centering
                \caption{Flickr}
                \label{fig:y}
        \end{subfigure}%
        \begin{subfigure}[b]{0.175\textwidth}
                \includegraphics[trim = 14mm 0mm 0mm 0mm, width=30mm]{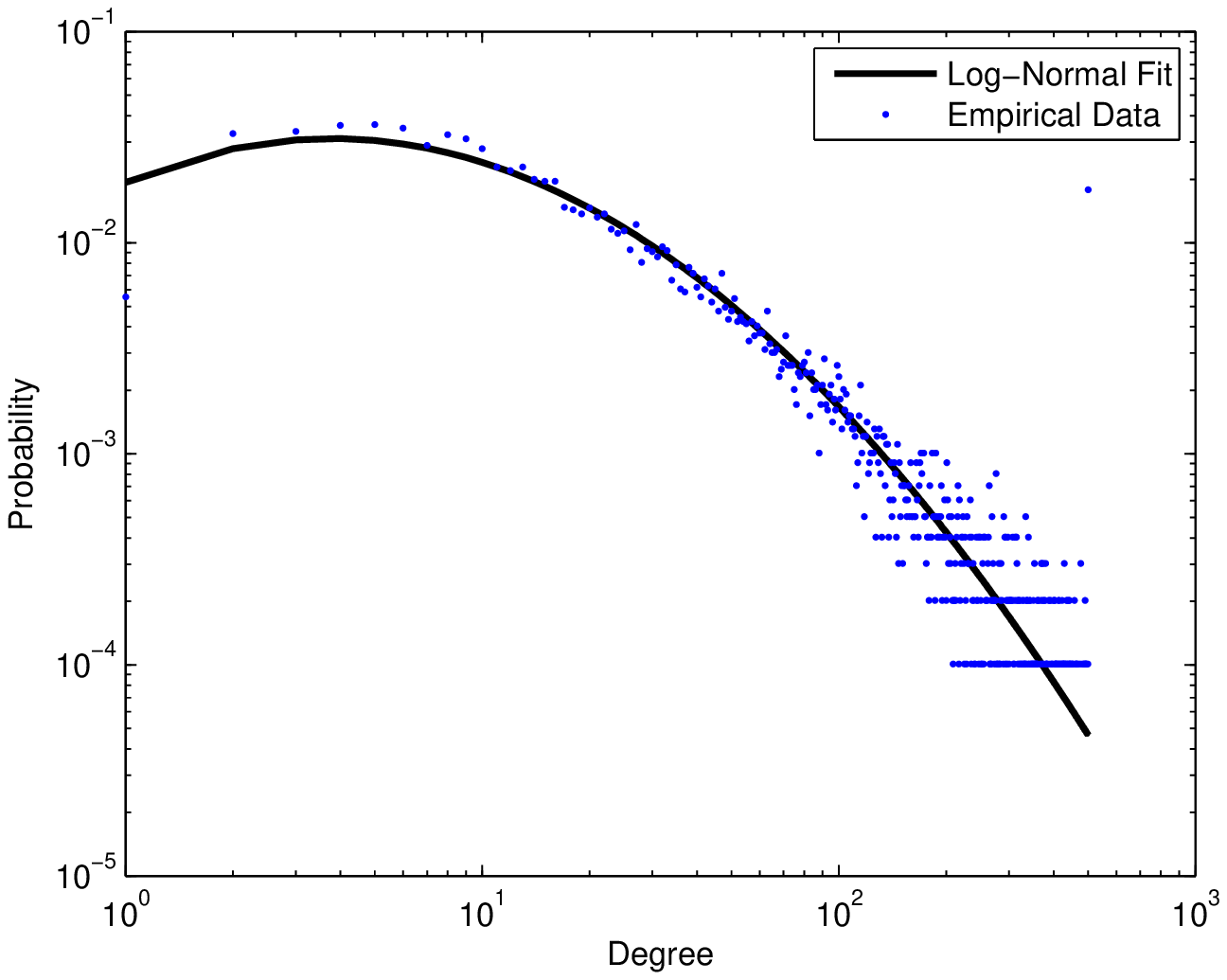}
                \centering
                \caption{Blog Cat.}
                \label{fig:z}
        \end{subfigure}%
        \caption{Empirical Node Degree Data and Fitted Log-Normal Probability Distribution Functions}\label{fig:animals}
\label{fig:22}
\end{figure}

As seen in Fig. \ref{fig:22}, all datasets follow a roughly Log-Normal distribution, with varying amounts of degree sparsity, and variance. Due to the high number of low degree nodes in the Google+ dataset, all points appear constrained to the left of the plot axis; however, as we will illustrate, the Log-Normal Distribution is still superior to the Pareto Distribution for link prediction.
During the training phase, $10\%$ of the data was removed in order to use for future predictions. For the purposes of demonstration, only $1,000$ of the highest degree nodes are maintained for adjacency matrix formation.

\section{Results}
\subsection{Baseline Methods and Performance Metrics}
In order to understand the advantage of LNMC, the results are compared against the following methods:

\begin{enumerate}
\item Matrix Completion with Pareto Sparsity (MCPS) - MCPS \cite{5} utilizes the same algorithm which we have outlined in the paper with the exception of the prior. MCPS employs the Pareto Distribution $f(d) = (\frac{\delta}{d})^\chi $.
\item Matrix Completion with $L_1$ Sparsity (MCLS) - MCLS is used by Richard et al. \cite{4}, and represents one of the first attempts at incorporating $L_1$ sparsity with the Low Rank assumption.
\item Logistic Regression (MF + RwR + AA) - In their paper on Social Attribute Networks, Gong et al. \cite{2} provide a method which combines features from Matrix Factorization, Random Walks with Restart, and Adamic Adar, which effectively solves the link prediction problem with high accuracy. In this paper, the attributes are removed from the network for equal comparison with our method.
\end{enumerate}

In order to provide a fair basis on which to judge the performance, Area Under the Curve (AUC) is employed for comparison. By utilizing the AUC as the performance metric, we avoid the need for data balancing, a process which frequently results in undersampling negative samples. Thus, all methods can benefit from the additional training data.
\\
The results are obtained via $10-\text{fold}$ Cross Validation, using a random sampling method for hyper-parameter selection. The rounds are averaged to produce the results shown in Table \ref{table:1}.
\subsection{Performance Comparison}
\begin{figure}[htb]
\centering
\includegraphics[width=90mm]{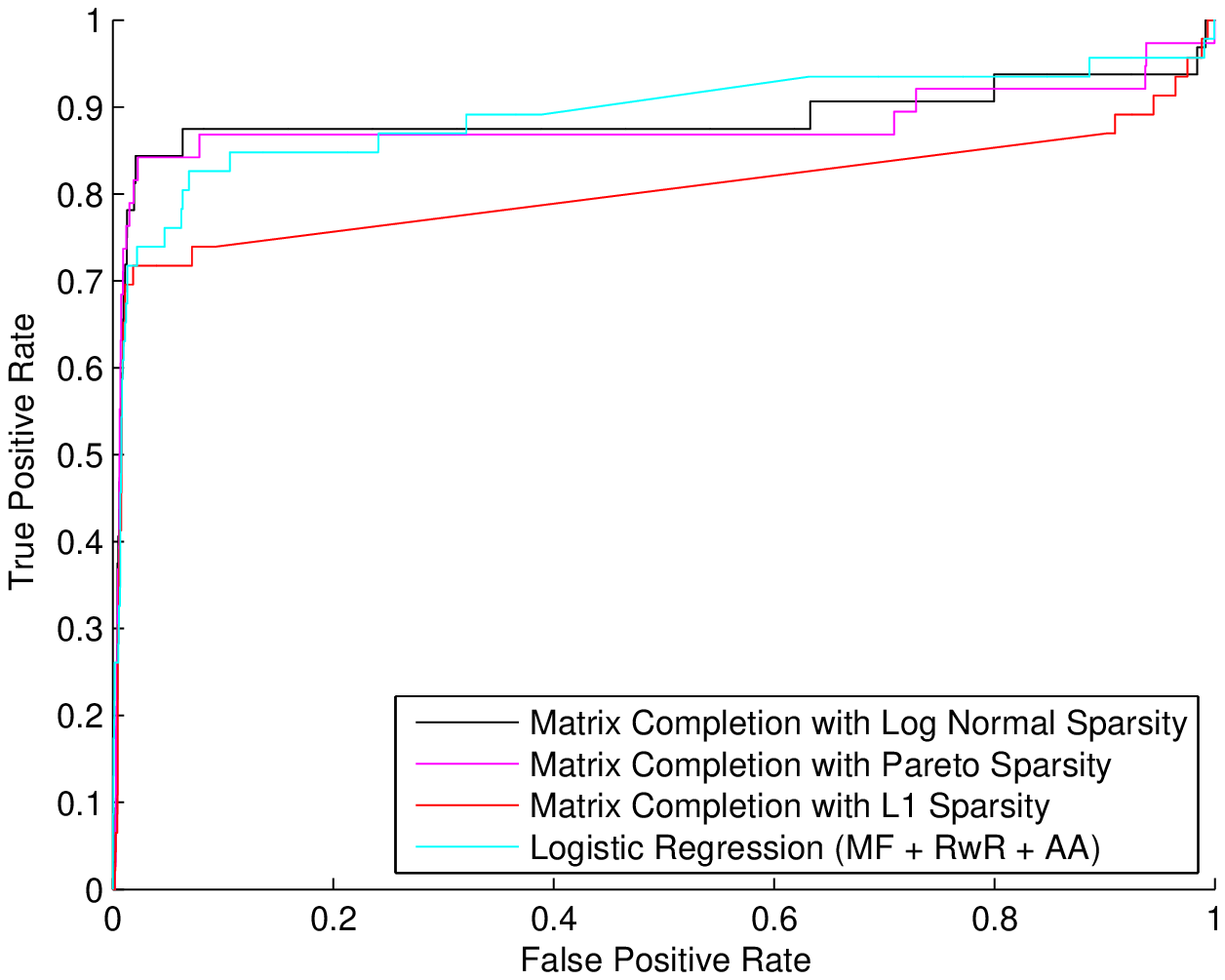}
\caption{Receiver Operating Characteristic for Google Plus Data}
\label{fig:1}
\end{figure}

\begin{figure}[htb]
\centering
\includegraphics[width=90mm]{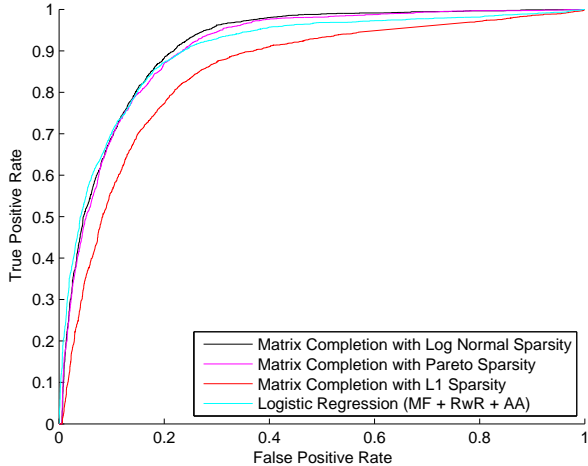}
\caption{Receiver Operating Characteristic for Flickr Data}
\label{fig:2}
\end{figure}

As demonstrated in Fig. \ref{fig:1}, LNMC outperforms MCPS, MCLS, and LR, on the Google Plus dataset. Due to the highly Log-Normal characteristic \cite{2} of the data set, LNMC's fine-tuned degree specific prior captures the degree distribution behavior in combination with the low rank features of the data, leading to high AUC values. The high number of true positives compared to the false positive rate leads to jagged graph distribution. In Fig \ref{fig:2}, it is clear that matrix completion with Pareto Sparsity produces low AUC values due to the inaccurate distribution representation. Similarly the LR method fails to capture accurate low rank information because the low rank matrix factorization is done prior to the the gradient descent training for Logistic Regression.
Due to the Pareto nature of the Flickr dataset, both the LNMC and MCPS methods perform the same. As can be seen in (\ref{eq:3}), LNMC can adapt to Scale Free Networks when the first term is small compared to the second term. Logistic Regression performs poorly since the features are set, whereas Matrix Completion methods automatically select the number of latent parameters to utilize.

\begin{figure}[htb]
\centering
\includegraphics[width=90mm]{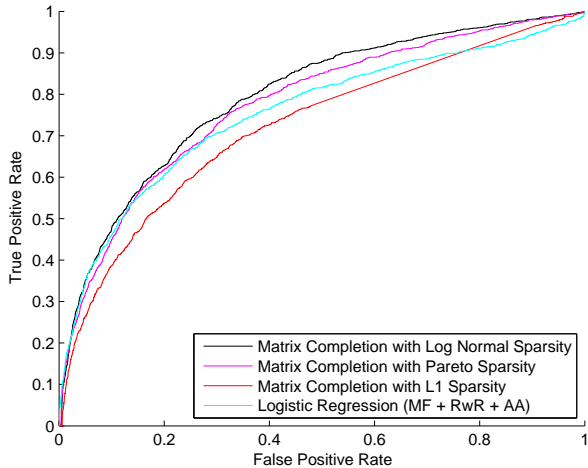}
\caption{Receiver Operating Characteristic for Blog Catalog Data}
\label{fig:3}
\end{figure}

As seen in Fig. \ref{fig:3}, LNMC outperforms the Pareto Sparisty based matrix completion, due to the inclusion of the squared log terms. The $L_1$ sparsity used in the MCLS method is insufficiently descriptive for accurate matrix estimation. Thus Logistic Regression, which incorporates more descriptive features outperforms the MCLS method.
For purposes of comparison, AUC values for each method and dataset, are contained in Table \ref{table:1}. As highlighted by the AUC Table, LNMC provides optimal results over all datasets.
\begin{table}[h]
\begin{center}
\begin{tabular}{|c|c|c|c|c|}
      \hline
      Data Set & LNMC & MCPS & MCLS & LR(MF+RwR+AA)\\
      \hline
      Google+ & .8541 & .8439 & .8113 & .8434\\
      \hline
      Flickr & .9052 & .9052 & .8504 & .8972\\
      \hline
      Blog Catalog & .7918 & .7846 & .7150 & .7727\\
      \hline
\end{tabular}
\vspace{.5cm}\\
\caption{AUC Performance Comparison}
\label{table:1}
\end{center}
\end{table}
\section{Conclusion}
As demonstrated both theoretically, and experimentally, LNMC is able to sufficiently encapsulate the advantages of Pareto Sparsity in addition to Log Normal Sparsity. Previously described by Gong et al. in \cite{2},
many modern social networks with undirected graph topologies exhibit Log Normal degree distributions. Thus by incorporating the degree-specific prior the optimization encourages convergence to a Log-Normal degree distribution.
Due to the non-convexity of solving the joint low-rank and structured sparsity inducing prior, the Lovasz Extension is introduced to solve the complex problem efficiently. Through analysis on three datasets, and using 3 top performing methods, we provide results which exceed the current optimum.
These results reveal the fundamental value of prior degree information in Link Prediction, and can provide insight into understanding the complex dynamics which cause links to form in a similar way across different networks.
\\
In future research we plan to investigate the incorporation of side information into the objective. Node attributes introduce additional challenges, including missing features, and additional training complexities.

\bibliographystyle{IEEEtran}
\bibliography{references}

\appendix
As seen in \cite{5}, the optimization of (\ref{eq:0}) is performed by first imposing the symmetry constraint on $Y$ as
\begin{gather*}
\argmin_Y \lambda_2 \Gamma(Y) + \frac{\mu}{2} \|X^{k+1} - Y + V^{k} \|_{2}^2 \\
\text{s.t. } Y=Y^{T}.
\end{gather*}
This minimization leads to the following algorithm:

\begin{algorithm}[!htb]
 \KwData{$X^{k+1}, V^{k},\mu, Yinit=(X^{k+1}+V^{k})$}
 \KwData{$\gamma, U=0_N,\omega$}
 \KwResult{$Y$}
 initialization\;
 \While{$\|Y-Y^T\|_2 < \omega$}{
 \For{ $r=0 \to N-1$}{
 $Y_{r,*} = \text{LovaszOptimize}(Yinit_{r,*},U_{r,*})$
 }
 $U=U+\gamma(Y-Y^T)$
 }
 $Y=\frac{1}{2}(Y+Y^T)$\\
 \Return{$Y$}
 \vspace{.5cm}\\
 \caption{Optimization with Symmetry Constraint}
\end{algorithm}

\begin{algorithm}[!htb]
 \KwData{$yinit, u, M$}
 \KwData{$d=yinit-u, p = 0_M$}
 \KwData{Set membership function $\zeta$}
 \KwData{$\theta$ transformation which translates sorted position index to original index}
 \KwResult{$y$}
 initialization\;
\For{$l=0 \to M-1$}
{
$q = \theta(l)$
$p_{q} = |d_q|- \frac{\lambda_2}{\mu}(\text{ln}^2 (l+1) - \text{ln}^2(l)+ \frac{(\sigma^2-\mu)(\text{ln}(l+1) - \text{ln}(l))}{\sigma^2})$
$\zeta(q).\text{value}=p_{q}$
$r=l$\\
\While{$r>1 \text{ and } \zeta(\theta(r)).\text{value} \geq \zeta(\theta(r-1)).\text{value}$}
{
$\text{Join the sets containing } \theta(r) \text{ and } \theta(r-1)$
$\zeta(\theta(r)).\text{value} = \frac{1}{|\zeta(\theta(r))}\sum_{i \epsilon \zeta(\theta(r))}p_i$
$\text{set: r to the first element of }\zeta(\theta(r))\text{ by sort ordering}$
}
}
\For{$j=1$ to \text{N}}
{
$y_j = \zeta(i).\text{value}$
\If {$y_j < 0$}
{
$y_q = 0$
}
\If {$d_i < 0$}
{
 $y_q = -y_q$
}
}
\Return{ $y$}
 \vspace{.5cm}
 \caption{LovaszOptimize Problem}
\end{algorithm}

\end{document}